\documentclass[12pt,preprint]{aastex}
\usepackage{epsfig}


\def\ls{{_<\atop^{\sim}}}

\lefthead{Nicastro et al.}

\begin{document}

\title{The Lack of BLR in Low Accretion Rate AGN as Evidence of
their Origin in the Accretion Disk}

\author{
Fabrizio Nicastro,$\!$\altaffilmark{1}
Andrea Martocchia,$\!$\altaffilmark{2,3}
Giorgio Matt$\!$\altaffilmark{2}
}

\altaffiltext{1}{Harvard-Smithsonian Center for Astrophysics, 
60 Garden st., Cambridge, MA 02138, USA, e-mail: fnicastro@cfa.harvard.edu}
\altaffiltext{2}{Dipartimento di Fisica, Universit\`a Roma Tre,
via della Vasca Navale 84, I-00146 Roma, Italy, e-mail: 
matt@haendel.fis.uniroma3.it}
\altaffiltext{3}{Observatoire Astronomique, 11 Rue de l'Universit\'e, 
F--67000 Strasbourg, France, e-mail: martok@isaac.u-strasbg.fr}


\begin{abstract}
In this paper we present evidence suggesting that the absence or 
presence of Hidden Broad Line Regions (HBLRs) in Seyfert 2 galaxies is 
regulated by the rate at which matter accretes onto a central 
supermassive black hole, in units of Eddington rate. 
Evidence is based on data from a subsample of type 2 AGNs extracted from 
the Tran (2001) spectropolarimetric sample and made up of all those sources 
that also have good quality X-ray spectra available, and for which a bulge 
luminosity can be estimated. 
We use the intrinsic (i.e. unabsorbed) X-ray luminosities of 
these sources and their black hole masses (estimated by using 
the well known relationship between nuclear mass and bulge luminosity in 
galaxies) to derive the nuclear accretion rate in units of Eddington. 
We find that virtually all HBLR sources have accretion rate larger 
than a threshold value of $\dot{m}_{thres} \simeq 10^{-3}$ (in Eddington 
units), while non-HBLR sources lie at $\dot{m} \ls \dot{m}_{thres}$. 

These data nicely fit predictions from a model proposed by Nicastro (2000), 
in which the Broad Line Regions (BLRs) are formed by accretion disk 
instabilities occurring in 
proximity of the critical radius at which the disk changes from gas pressure 
dominated to radiation pressure dominated. This radius diminishes with 
decreasing $\dot{m}$; for low enough accretion rates (and therefore 
luminosities), the critical radius becomes smaller than the innermost stable 
orbit, and BLRs cannot form. 
\end{abstract}


\section{Introduction}
Seyfert 2 galaxies (where only narrow emission lines are 
visible) are commonly believed to be intrinsically the same 
as Seyfert 1 galaxies (where both narrow and broad emission 
lines are visible), the difference being due to orientation. 
According to the widely accepted Unification Model for AGNs 
(Antonucci, 1993) type 2 AGNs are seen edge-on, through large 
columns of circumnuclear obscuring material that prevents 
the direct view of the nucleus, including the Broad Line
Regions (BLRs). 
This scenario was first proposed by Antonucci \& Miller 
(1985) to explain the presence of polarized broad lines in the archetypical  
Seyfert 2, NGC 1068, and is now supported other than by spectropolarimetric 
observations of Hidden Broad Line Regions (HBLRs) in several other sources, 
also by X--ray observations, which demonstrate that Seyfert 2s have usually 
absorption columns largely exceeding the Galactic ones. 

Despite observations do generally support orientation based 
unification models for AGNs, exceptions do exist. Only about 
50 \% of the brightest Seyfert 2s show the presence of HBLRs (Tran, 2001) 
in their polarized optical spectra, while the remaining half do not. 
It has now been convincingly shown that the presence or absence of 
HBLRs in Seyfert 2s depends on the AGN luminosity, with the HBLR 
sources having on average larger luminosities (Lumsden \& Alexander
2001; Gu \& Huang 2002; Martocchia \& Matt 2002; Tran 2001, 2003). 
While Lumsden \& Alexander (2001) explained this finding still in the
framework of an orientation model, Tran (2001) proposed the 
existence of a population of galactic nuclei whose activity 
is powered by starburst rather than accretion onto a supermassive black 
hole and in which therefore the BLRs simply do not exist. 

In this paper we present evidence that suggests that the absence or 
presence of HBLRs is regulated by the ratio between the X-ray luminosity 
and the Eddington luminosity which, in the accretion--power scenario, is a 
measure of the rate at which matter accretes onto the central supermassive 
black hole.
Our explanation is based on the model proposed by Nicastro (2000, 
hereinafter N00), in which the BLRs are formed by accretion disk 
instabilities occurring in proximity of the critical radius at which 
the disk changes from gas pressure dominated to radiation pressure 
dominated. This radius diminishes with decreasing $\dot{m}$; for low 
enough accretion rates (and therefore luminosities), the critical radius 
becomes smaller than the innermost stable orbit, and BLRs cannot form. 
Under the Keplerian assumption, the model naturally predicts that 
AGNs that are accreting close to the lowest possible $\dot{m}$ 
must show the broadest possible emission lines in their optical spectra, 
either hidden (i.e. in polarimetric light), if the nucleus is obscured, or 
not, if the nucleus is not obscured. 
An analogous model has been proposed recently by Laor (2003). In both 
Nicastro and Laor's models the existence of BLRs in AGN is related to the 
breadth of the broad emission lines (BELs), and based on the observed upper 
limit of $FWHM \sim 25000$ for the BELs. 
However, while in Nicastro's (2000) model this a consequence of the lines 
been produced in clouds of gas located at a distance from the source 
depending on the accretion rate in Eddington units (so the accretion rate 
is the physical driver of the observed correlation), in Laor's model the 
driving parameter is the width of the line itself, and no physical origin is 
proposed (other than the observed correlation between AGN luminosity 
and line width).

To check this hypothesis and test the above models, we extracted from 
the Tran (2001) spectropolarimetric sample, the subsample of type 2 AGNs that 
also have good quality X-ray spectra available (see Martocchia \& Matt 2002). 
Most of the sources in our sample have [O {\sc iii}] luminosities much 
lower than $10^{43}$ erg s$^{-1}$. At such relatively low luminosities, 
stellar light from the galaxy may be a strong contaminant in the optical 
band. X--ray luminosities, instead, are little or not contaminated by 
stellar components and so are possibly more reliable indicators of the 
nuclear, accretion-powered activity. 
For each source of our sample, we then estimate its 2-10 keV 
intrinsic (i.e. unabsorbed) luminosity, and use it as a reliable measure 
of the nuclear activity. 
We then estimate the mass of the black hole, using the relation between the 
mass and the bulge luminosity (Magorrian et al. 1998; Ferrarese \& 
Merrit 2000). Comparing the black hole mass and the X-ray luminosity, 
the accretion rate is eventually derived.  
 
\bigskip
The paper is organized as follows: in Sec. 2 we define and discuss the 
adopted sample and the methods used to derive the various parameters. 
In Sec. 3 the results of our analysis are presented, and discussed in 
Sec. 4. 

Throughout this paper a value of $H_0$=70 km s$^{-1}$ Mpc$^{-1}$ is
adopted. 

\section{The sample}
We extracted our primary sample from the spectropolarimetric survey of 
Tran (2001; see also Tran 2003) of all, optically classified, Seyfert 2 
galaxies in the CfA and 12$\mu$ samples. Optical polarized spectra 
of the sources in this survey are rather homogeneous in signal-to-noise 
(Tran, 2001), 
so the lack of polarized Broad Lines in the optical spectra of about 
half of the sources in the sample should not be an artifact of dramatic 
differences in data quality. 
First we searched for the X-ray properties of the sample [namely (a) the 
intrinsic 2-10 keV flux, and (b) the equivalent Hydrogen column 
density $N_H$], and selected only those sources that had been observed at 
least once with imaging X-ray satellites. This allowed us to minimize 
confusion problems which, at the typical flux level of our sample, may 
be relevant (Georgantopoulos \& Zezas, 2003). 
In practice, we used data only from ASCA, {\it Beppo}SAX, {\it Chandra} and
XMM--{\it Newton}, either taken from literature, when available and 
reliable, or (re)analyzed by ourselves. 
Finally, we applied two further selections, discarding (a) all 
sources suspected to be Compton--thick, and (b) sources known to be 
strongly variable, in X-rays, on time scales of years (e.g. NGC 7172, 
Dadina et al. 2001). For Compton--thick sources no direct measurements 
of the intrinsic luminosity is available, while highly variable sources 
cannot be assigned univocally a X-ray luminosity.
Additional details on our primary sample selection can be found in 
Martocchia \& Matt (2002).
We ended up with a small but reliable sample consisting of 10
HBLR and 6 non-HBLR sources. 
Previous works (Gu \& Huang, 2002, Martocchia \& Matt, 2002, 
Tran, 2003) have shown that all HBLR sources have 2-10 keV luminosities 
larger than $\sim3\times10^{42}$ erg s$^{-1}$, while all non-HBLR sources 
have luminosities smaller than this value. 
No correlation with the column density of the absorber is, instead, found. 
The X--ray luminosity is clearly correlated with the [O {\sc iii}] 
luminosity (even if with a large scatter), confirming the reliability of 
the estimates.

\bigskip
For the 16 sources in our primary sample, we then tried to estimate 
the nuclear accretion rate, in units of Eddington. To calculate $\dot{m}$, 
we needed to evaluate the mass of the central black hole, which we 
derived using the Ferrarese \& Merritt (2000) empirical relationship 
between nuclear mass and bulge luminosity (we used the numerical coefficients 
they derived from their sample A). The bulge luminosity $L_{bulge}$ was 
derived from the empirical correlation between the galaxy's morphological 
type (i.e. T-type) and the bulge to total luminosity ratio given by Simien 
\& de Vaucouleurs (1986) 
\footnote{
The $M_{BH}$-$\sigma$ correlation is more tight than the 
$M_{BH}$-$L_{bulge}$ correlation, as discussed in detail
by Ferrarese \& Merritt (2000). However, for only 4 sources (three of
them HBLR) in our sample we could find measured nuclear velocity dispersions
(only 2 more can be added, both HBLR, considering all sources in the 
Gu \& Huang 2002 compilation). 
Moreover, sometimes the reported values differ significantly, with 
dramatic effects on the estimate of the mass, even of an order of magnitude. 
For these reasons we were forced to use the $M_{BH}$-$L_{bulge}$ 
relationship, despite its larger scattering.} 
. The (corrected for extinction) total luminosity and T-type were taken from 
the RC3 catalog (de Vaucouleurs et al. 1991). 
The complete information was available only for a subset of our primary 
sample. This further selection reduced then our final primary sample to a 
total of 10 sources: 6 HBLRs and 4 non-HBLRs. 

For completeness, and given the limited size of our final primary sample, 
we also considered a secondary sample, derived applying the same criteria 
and rules used to select our primary sample to the  
Gu \& Huang (2002) compilation. 
This compilation contains all Seyfert 2s with published (between 1995 and 
2002) spectropolarimetric information, and therefore is not as homogeneous 
as the Tran (2001) sample in the quality of the polarized optical spectra. 
This search added 3 more HBLRs and 2 more non-HBLRs to our final 
sample. 

For all sources in our final sample we calculated the accretion rate in 
Eddington units (actually the ratio between the nuclear bolometric 
luminosity $L$ and the Eddington luminosity $L_E$) assuming a factor 10 
correction between the 2-10 keV and bolometric luminosities (see footnote 5 
in \S 3). 
Results are summarized in Table~\ref{lmdot} (for the sources of our 
secondary sample luminosities are slightly different from those reported 
by Gu \& Huang, 2002, because of the different choice of H$_0$). 

%
\begin{table}
\begin{center}

\vspace{0.4truecm}
\begin{tabular}{|c|ccccc|}
\hline
Source name & $L_X$ & M$_{bh}$ & $\dot{m}$ & HBLR & Sample \cr
& 10$^{42}$ erg s$^{-1}$ & 10$^{7}$ M$_{\odot}$ & 10$^{-3} $L/L$_E$ & & \cr
\hline
\hline
& & & & & \cr
IC 5063 & 8.1 & 28 & 2.2 & Y & P \cr 
MKN 438 & 10.5& 12 & 6.6 & Y & P \cr
NGC 4388 &  6.4  &  35 & 1.4 & Y & P \cr
NGC 5506 &  8.7 & 9.9 & 6.6 & Y & P \cr
NGC 6552 & 3.1  & 15 & 1.5 & Y & P \cr
MKN 1210 & 16.3  & 10 & 13 & Y & P \cr
NGC 3081 & 0.67 & 13 & 0.59 & Y & S \cr
NGC 4507 & 16.5 & 18 & 8.1 & Y & S \cr
NGC 5252 & 11.6 & 36 & 2.8 & Y & S \cr
& & & & & \cr
\hline
& & & & & \cr
M51  & 0.007 & 6.8 & 0.009 & N & P \cr
NGC 3079 & 0.018 & 1.4 & 0.09 & N & P \cr
NGC 4941 & 0.086 & 5.1 & 0.13 & N & P \cr
NGC 7582 & 1.57 & 20 & 0.6 & N & P \cr
NGC 3281 & 6.28 & 28 & 2.0 & N & S \cr 
NGC 7590 & 0.06 & 6.3 & 0.084 & N & S \cr 
& & & & \cr
\hline
\end{tabular}
\end{center}
\caption{2-10 keV X--ray luminosities, black hole masses and accretion
rates (defined as the ratio of bolometric to Eddington luminosity) for our 
final Primary (P) and Secondary (S) samples.}
\label{lmdot}
\end{table}
%

\section{Results}
It has been recently shown (Gu \& Huang, 2002, Martocchia \& Matt, 2002, 
Tran, 2003) that the intrinsic AGN luminosity 
of the sources of the Tran (2001) sample can be used
to clearly separate the two classes of HBLR and non-HBLR sources. 
Sources with 2-10 keV luminosities larger than the threshold value of 
$L_X^{thres} \simeq 3 \times 10^{42}$ erg s$^{-1}$ do show HBLs in their 
optical polarized spectra, while sources with $L_X < L_X^{thres}$ do not 
(see Figure 1 of Martocchia \& Matt, 2002). 
Here we provide evidence that suggests that this separation is due to 
differences in nuclear accretion rate, from HBLR to non-HBLR sources. 

Figure 1 shows fractional luminosities in units of Eddington 
luminosities, versus black hole masses for all the 15 sources of our 
primary (circles) and secondary (squares) samples. Open symbols in this 
plot represent HBLR sources, while filled symbols represent non-HBLR 
sources. We first note that a very broad range of accretion rates is 
spanned by the sources of our sample (more than three orders of magnitude), 
which are otherwise powered by central black holes with rather homogeneous 
masses (only a factor of about 15 across the entire sample). 
Most importantly, Figure 1 clearly shows that HBLR sources are accreting 
at much faster rates compared to non-HBLR sources. 
\footnote{We note that this result is independent on the particular value 
chosen to convert 2-10 keV into bolometric luminosities: a different value 
would only shift the threshold accretion rate value that separates 
HBLR from non-HBLR sources. Of course a potentially more serious problem 
may arise if the sources in our sample have Spectral Energy Distributions 
that dramatically differ from each other. 
However, such a random effect would likely destroy rather than create 
the correlation that we find.}
.

\noindent
The threshold value of $\dot{m}_{thres} \simeq 10^{-3}$ divides up HBLR from 
non-HBLR sources in the $M_{BH}$ vs $\dot{m}$ plane (dashed vertical line 
in Fig. 1). 
The only exceptions are NGC~3081 and NGC~3281, both sources from our 
``secondary'' sample. NGC~3081 is a HBLR source, with an accretion rate 
of $\sim 6 \times 10^{-4}$, a factor of $\sim 2$ below the threshold 
value, while NGC~3281 is a non-HBLR source accreting at a rate of 
$\sim 2 \times 10^{-3}$, again a factor of 2, but above the threshold. 
We looked at these two exceptions in somewhat higher detail.
The flux of NGC~3081 reported in Gu \& Huang (2002) is taken from Maiolino 
et al. (1998; BeppoSAX data). We reanalyzed those data and found very 
similar results. However, the spectral
parameters are very poorly constrained. If the power law index is
fixed to 2 (instead of the value of 1.7 chosen by Maiolino et
al. 1998), an almost equally good fit is found, but the inferred luminosity 
(and then $\dot{m}$) is twice as large 
\footnote{This large uncertainty is due to the low S/N of the observation
coupled with the large value of the $N_H$.}
, moving the source to the right side of our $\dot{m}$ vs $ M_{BH}$ 
plane, and so supporting our main conclusion. 

\noindent 
The X--ray luminosity reported by Gu \& Huang (2002) for the second 
exception in our sample, the non-HBLR source NGC~3281, is based on an 
ASCA observation (Bassani et al. 1999), and is underestimated by a factor 
$\sim$3. A subsequent BeppoSAX observation, in fact, demonstrated that the 
source is actually moderately Compton--thick (Vignali \& Comastri, 2002), 
which implies a higher intrinsic luminosity. This, in turn, implies a higher 
accretion rate compared to that reported in Table 1, and then amplify, 
rather than moderate, the discrepancy found. 
However, for this source we could not find in the literature
details on the spectropolarimetric observation, so it is impossible
to judge how significant is the upper limit on the presence of polarized 
broad lines. 

%
\begin{figure}
\plotone{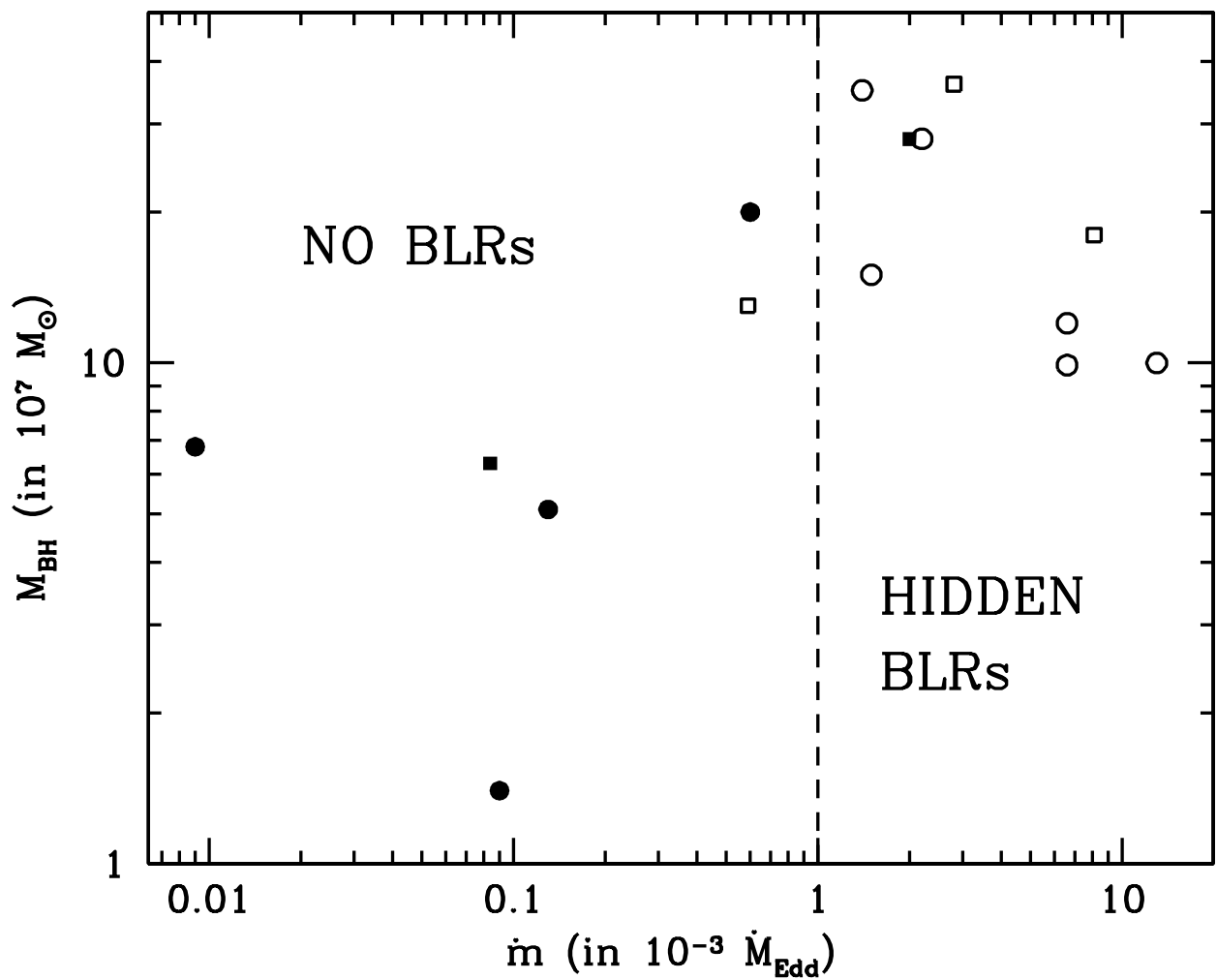} 
\vspace{0in}
\caption[h]{Black hole masses vs accretion rates (defined as $\dot{m} = 
\dot{M}/\dot{M_{Edd}} = L_{bol}/L_{Edd}$). Open and filled symbols 
refer to HBLRs and non-HBLRs. Circles and squares are sources from our 
primary and secondary samples, respectively.}
\label{m_mdot}
\end{figure}
%

\section{Discussion}
We have presented evidence for a correlation between the presence
of HBLs in the polarized optical spectra of nearby Seyfert 2s, and their 
nuclear accretion rate. 
Virtually all sources with HBLRs are found above the threshold value of 
$\dot{m} = 10^{-3}$, in the $\dot{m}$ vs $ M_{BH}$ plane, and viceversa. 
Our sample is admittedly small, but both the spectropolarimetric 
observations (at least for the sources of our primary sample) and the 
X--ray luminosities are quite reliable. 
It should be also noted that the many sources of uncertainties in our 
$\dot{m}$ estimates [namely (a) the bulge-to-total luminosity ratio vs 
T type, (b) the black hole mass to bulge luminosity correlation, and (c) 
the bolometric correction], would likely destroy, certainly not artificially 
create, the observed correlation. 

Our findings fit nicely with predictions of the model of N00, 
in which the BLRs originate from the accretion disk at the transition
radius between the gas pressure and radiation pressure dominated
regions. This suggests that the accretion rate (N00), rather than the 
``line-width'' (Laor, 2003) is indeed the parameter physically responsible 
for the presence or not of BLRs in AGN.

\noindent
It is remarkable that the threshold value of $\dot{m}_{thres} 
\simeq 10^{-3}$ that we find is so close to what predicted in the framework 
of the N00 model (i.e. about 4$\times10^{-3}$ for a black hole mass of 
10$^8$M$_{\odot}$). 
The threshold value $\dot{m}_{thres}$, however, is almost independent of 
the spin of the black hole, and so, unfortunately, this model cannot be 
used to discriminate between spinning and non-spinning black holes. 
This is because in the N00 model the critical radius at which the BLR 
should form decreases with increasing radiative efficiency in the disk 
and this decrease is almost exactly balanced by the decrease of the radius 
of the last stable orbit. 

\bigskip
Of course, if our explanation is correct, a fraction (set by the relative 
number of type 1 to type 2 Seyferts in the nearby Universe) of the non-HBLRs 
should be actually not obscured. 
In the Gu \& Huang (2002) compilation, 4 out of the 23 non-HBLR 
sources with X-ray information available (i.e. 17 \%) show little 
($N_H < 10^{21}$ cm$^{-2}$) cold X-ray absorption. In the 
Martocchia \& Matt (2002) sample (i.e. our primary sample, before the bulge 
luminosity estimate selection) 1 out of the 6 non-HBLR sources (again, 17 \%) 
has an estimated column density only a factor of about 3 larger than 
the Galactic column along that line of sight. 
In our final sample only one object (NGC~7590, extracted from the Gu \& 
Huang, 2002, compilation) is virtually unabsorbed. 
Our estimate of the accretion rate for this objects could then 
in principle be affected by the AGN light in this source, which 
would artificially increase the total galaxy luminosity (the observed 
parameter we use), and so the bulge one. 
However, the 2-10 keV luminosity of this source is very low (see table 1) 
and so presumably is its B band luminosity. The effect on the estimate of 
$\dot{m}$, in this case, should therefore be small. 

Though suggestive of the existence of a population of unobscured (i.e. 
type 1) non-HBLR sources, the current statistics is too poor to draw 
any firm conclusion. 
Moreover current estimates of equivalent H column densities are based on 
X-ray observations taken with satellites with poor spatial resolutions 
(mostly ROSAT, ASCA and {\em Beppo}SAX). At such low luminosities, 
a single spatially unresolved luminous X-ray binary, or a population of 
those, could lead to underestimate the column of matter actually 
absorbing the nuclear light (Georgantopoulos, \& Zezas, 2003). 
Better (i.e. higher spatial resolution and signal-to-noise) and more 
numerous X-ray data are required to disentangle this issue. 
Enlarging the sample of low luminosity supposedly type 2 AGNs with both 
good quality optical spectropolarimetric and X-ray (i.e. {\em Chandra} 
and/or XMM--{\em Newton}) data, would allow us to eventually find 
a sub-population of non-HBLR sources, with none or little cold X-ray 
absorption. These would be the genuine type 1 counterparts of the low 
accretion rate non-HBLR ``true'' type 2 sources of our current sample. 
This population should fall on the left side ($\dot{m} < \dot{m}_{thres}$) 
of the $\dot{m}$ vs $M_{BH}$ plane (see also Laor, 2003). 

We note that, recently, few examples of this class of low-accretion rate 
type 1 AGNs may have actually been discovered in X-rays (i.e. Pappa et al., 
2001, Georgantopoulos, \& Zezas, 2003, Boller et al., 2002). 
The ASCA and {\em Chandra} spectra of the optically classified Seyfert 2 
NGC~4698, shows little or no absorption (Pappa et al., 2001, 
Georgantopoulos, \& Zezas, 2003). The estimated accretion rate for the 
central AGN in this galaxy is lower than $\dot{m} \ls 10^{-4}$ 
(Georgantopoulos, \& Zezas, 2003), consistent with our proposed scenario. 
Similarly ROSAT and 
{\em Chandra} spectra of 1ES~1927+654 shows an unobscured, highly 
variable steep power law continuum, typical of Narrow Line Seyfert 1s 
(Boller et al., 2002). This object has been recently re-classified as a 
Seyfert 2 based on its optical spectrum (Bauer et al., 2000). 1ES~1927+654 
has an optical (B-band) luminosity of $\sim 10^{43}$ erg s${-1}$, but 
no mass estimate is available for its central black hole, and so for 
the accretion rate in Eddington units. However, given the overall X-ray 
properties of this object (i.e. steep spectrum and large amplitude 
variability), we speculate that 1ES~1927+654 is actually accreting 
at very high rates, in terms of Eddington, and belongs then to the opposite 
extreme of the line-width vs accretion rate correlation in the framework of 
the N00 model. If this is the case, 1ES~1927+654 would actually be an 
optically mis-identified Narrow Line Seyfert 1. 
The two objects described above have no spectropolarimetric observation, 
so we do not know whether they have a HBLR or not.

\bigskip 
Our findings are based on X-ray luminosities and black hole masses derived 
from the mass-to-bulge luminosity relation. It is interesting to note that 
Gu \& Huang (2002), also find a similar accretion-rate versus 
HBLR-existence correlation (see their Fig. 6), using [O {\sc iii}] (rather 
than X-ray) luminosities and black hole masses obtained from the 
M$_{BH}$-$\sigma$ correlation. 

\bigskip 
The results presented here support the N00 model, and suggest therefore 
that the accretion rate is the physical driver of the observed bimodal 
distribution of HBLR and non-HBLR sources with luminosities. 
While other explanations (see e.g. Lumsden \& Alexander 2001; Tran 2001, 
2003, Martocchia \& Matt, 2002, Gu \& Huang, 2002) cannot at present be 
ruled out, the results presented here are very encouraging, and worth 
checking more throughly by enlarging the sample with new spectropolarimetric 
and/or X--ray observations.

\begin{acknowledgements} 
Part of this work has been done during a visit of GM at CfA, whose 
hospitality he gratefully acknowledges. 
We thanks Andreas Zezas for enlightening discussions, and an anonymous 
referee for useful comments that helped improved the paper. We acknowledge
financial support from {\em Chandra} grant GO2-3122A (FN), MIUR under 
grant {\sc cofin-00-02-36} (AM and GM), and CNES (AM). 
\end{acknowledgements}


\end{document}